# *proFIA*: A data preprocessing workflow for Flow Injection Analysis coupled to High-Resolution Mass Spectrometry

Alexis Delabrière[1,*], Ulli M. Hohenester[2], Benoit Colsch[2], Christophe Junot[2], François Fenaille[2], and Etienne A. Thévenot[1,*]

[1]CEA, LIST, Laboratory for data analysis and systems' intelligence, MetaboHUB, F-91191 Gif-sur-Yvette, France
[2]CEA, DRF/JOLIOT/SPI, Laboratoire d'Etude du Métabolisme des Médicaments, MetaboHUB, F-91191 Gif-sur-Yvette, France

[*]To whom correspondence should be addressed.

## Abstract

**Motivation:** Flow Injection Analysis coupled to High-Resolution Mass Spectrometry (FIA-HRMS) is a promising approach for high-throughput metabolomics. FIA-HRMS data, however, cannot be preprocessed with current software tools which rely on liquid chromatography separation, or handle low resolution data only.
**Results:** We thus developed the *proFIA* package, which implements a suite of innovative algorithms to preprocess FIA-HRMS raw files, and generates the table of peak intensities. The workflow consists of 3 steps: i) noise estimation, peak detection and quantification, ii) peak grouping across samples, and iii) missing value imputation. In addition, we have implemented a new indicator to quantify the potential alteration of the feature peak shape due to matrix effect. The preprocessing is fast (less than 15 s per file), and the value of the main parameters (*ppm* and *dmz*) can be easily inferred from the mass resolution of the instrument. Application to two metabolomics datasets (including spiked serum samples) showed high precision (96%) and recall (98%) compared with manual integration. These results demonstrate that *proFIA* achieves very efficient and robust detection and quantification of FIA-HRMS data, and opens new opportunities for high-throughput phenotyping.
**Availability:** The *proFIA* software (as well as the *plasFIA* data set) is available as an R package on the Bioconductor repository (http://bioconductor.org/packages/proFIA), and as a Galaxy module on the Main Toolshed (https://toolshed.g2.bx.psu.edu/) and on the Workflow4Metabolomics online infrastructure (http://workflow4metabolomics.org).
**Contacts:** alexis.delabriere@cea.fr and etienne.thevenot@cea.fr.
**Supplementary information:** Supplementary data are available at *Bioinformatics* online.

# Introduction

Metabolomics is the comprehensive detection, quantification, and identification of the small molecules (< 1 kDa) involved in, or produced by, the metabolism (Fiehn, 2002). As a major approach of functional genomics and systems biology (Nicholson and Lindon, 2008), metabolomics provides new insights into physiology and disease (Zhang and Watson, 2015; Wishart et al., 2016).

Mass spectrometry (MS) is a key technology for biomarker discovery, due to its high sensitivity and large dynamic range (Dettmer *et al.*, 2007; Boudah *et al.*, 2013). Coupling of Liquid (or Gas) Chromatography to MS (LC- or GC-MS), or Capillary Electrophoresis (CE-MS) is often used to separate chemical species, and reduce the matrix effect (decrease of intensity affecting some of the ions, which may be due to charge competition during the ionization process, and increases with sample concentration or complexity; Enke, 1997; Taylor, 2005, Trufelli *et al.*, 2011). While such hyphenated techniques provide extensive coverage and characterization of the metabolome (Roux *et al.*, 2012), the chromatographic step is time-consuming (e.g., 30 min per sample for LC-MS). Therefore, large-scale phenotyping calls for the development of complementary very high-throughput approaches.

Flow Injection Analysis (FIA) consists in the transient injection of the sample into a carrier stream which flows directly into the MS instrument, without any chromatographic system (Hansen, 2008). Initial coupling of FIA to low resolution MS allowed successful sample classification based on nominal metabolite fingerprints (Allen *et al.*, 2003). Due to the absence of chromatographic separation, FIA is less selective than hyphenated approaches, because the retention time cannot be used to distinguish between isomers. In addition, the simultaneous introduction of all the analytes into the mass spectrometer results in stronger matrix effects and ion suppression. Evaluation of such ion suppression effects (mainly for minor metabolite species) constitutes a prerequisite to accurate FIA experiments. Despite these limitations, the coupling of FIA to modern high-resolution instruments (FIA-HRMS; mass accuracy typically around 1 ppm), which enables the efficient discrimination between isobaric compounds and facilitates the determination of molecular formulas, has proved to be a key technology for high-throughput metabolomics analysis (Madalinski *et al.*, 2008; Fuhrer *et al.*, 2011; Draper *et al.*, 2013).

Feature extraction in mass spectrometry is achieved by detecting and quantifying peaks in the raw files, followed by alignment (or grouping) between samples to generate a single table of feature by sample intensities. It is a critical step in MS data analysis to avoid false positives and achieve accurate quantification (Castillo *et al.*, 2011). Since MS data are complex (e.g. combining mass and time measurements), noisy, and of high-volume, the development of robust and efficient algorithms is challenging (America and Cordewener, 2008).

Several open-source software tools exist for LC-MS preprocessing, such as XCMS (Smith *et al.*, 2006), MZmine (Katajamaa *et al.*, 2006), and OpenMS (Sturm *et al.*, 2008). Such LC-MS software algorithms, however, cannot be used to preprocess FIA-MS data because of the high variability of the peak shapes in the time dimension for FIA acquisitions compared with liquid chromatography. For example, a strong matrix effect affecting the central part of a FIA peak may result in the detection of two peaks by the *centWave* algorithm. Specific preprocessing of FIA-MS data has been implemented by Enot *et al.* (2008) in the *FIEmspro* R package. The algorithm, however, has been developed for low resolution MS and relies on nominal mass binning, which does not take advantage from the high-resolution of recent mass spectrometers.

We therefore developed an innovative suite of algorithms for accurate and robust preprocessing of FIA-HRMS data. In the following sections, we present the underlying approach and modeling (*Approach*), the successive algorithms of the workflow (*Methods*), and the application of our software to a metabolomics data set (*Results*).

## Approach

Flow Injection Analysis coupled to Mass Spectrometry (FIA-MS) is characterized by the transient injection of the sample into a continuous solvent carrier directly connected to the Electrospray Ionization Source (ESI) of the mass spectrometer (Trojanowicz and Kolacinska, 2016). Each FIA-MS data file consists of successive acquisitions of mass spectra within a time window spanning the introduction of the sample (typically a few seconds or minutes), and has a classical three-dimensional structure (m/z, time and intensity). The first preprocessing step therefore consists in determining the m/z slices containing ion signal (*Band detection*). The second step, analyte quantification, involves several tasks in the time dimension: detection of the signal time limits, quantification of the signal intensity, and discrimination from the baseline solvent. Due to the specific mode of sample introduction into the mass spectrometer, elution profiles from the detected compounds in FIA strongly differ from direct infusion (DI-MS) and liquid chromatography (LC-MS) analysis, and therefore require dedicated signal processing algorithms (Fig. 1). By analogy with the term "chromatogram" used in LC- and GC-MS, elution profiles in FIA-MS will thereafter be referred as "flowgrams" (Ruzicka and Christian, 1990).

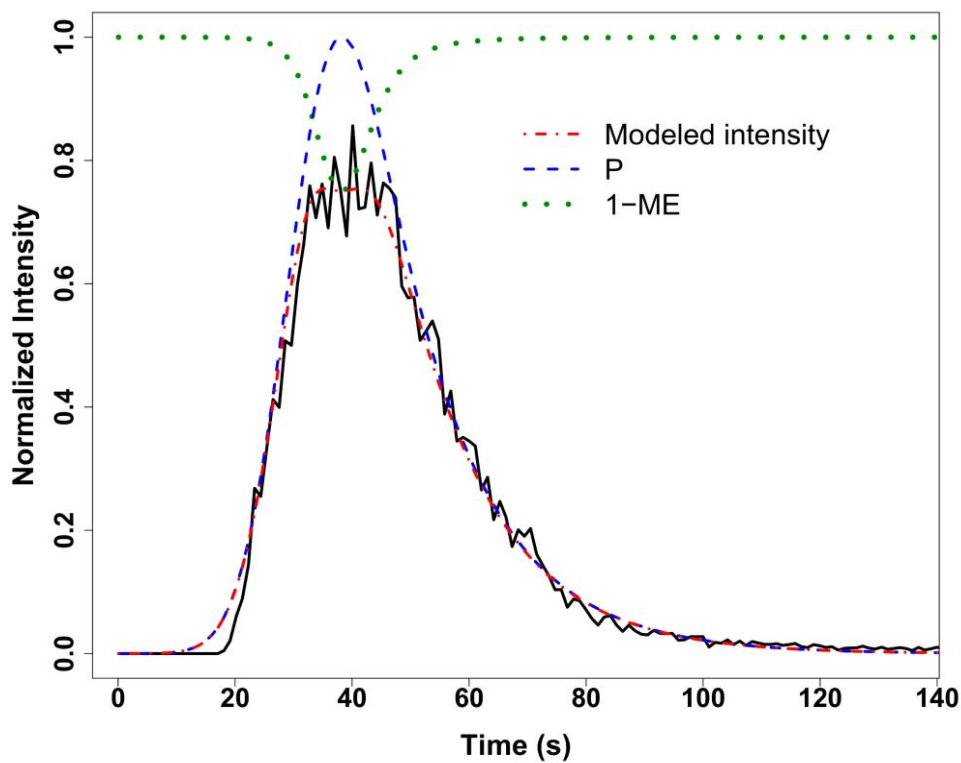

**Fig. 1. proFIA approach for peak detection and quantification.** An FIA flowgram (solid black line) can be modeled (dash-dotted red line) as the sum of four components (Equation 1): a signal proportional to the sample peak (Exponentially Modified Gaussian; dashed blue line), the matrix effect (exponential function of the sample peak; dotted green line; here $1 - ME$ is plotted for visualization purpose), the solvent baseline (not shown), and a heteroscedastic noise.

In the case of a sample containing a single analyte $A$, the expected flowgram, $P_A(t)$, would contain a single peak (since the sample is transiently introduced into the carrier flow, contrary to continuous infusion in DI-MS). This model can also be written $k_A \times P(t)$, where $k_A$ is a (constant) response factor depending of the ionization properties of the analyte (see Table 1 detailing the notations used in this section), and $P(t)$ is the normalized sample profile which depends on the flow injection conditions only (throughput, time window, fluid properties). Models for the peak $P$ (Kolev, 1994), such as the Exponentially Modified Gaussian, are usually right-tailed to take into account the dispersion of the sample zone into the carrier by convection and diffusion (Fig. 1).

In the case of a biological sample, all analytes enter the instrument at the same time (as there is no chromatographic separation, contrary to LC-MS and GC-MS). $P(t)$ can therefore still be used to model the global introduction of the sample into the mass spectrometer. Within each analyte flowgram, however, the matrix effect potentially affects the peak shape and must be taken into account. Matrix effect is defined as "the combined effect of all components of the sample other than the analyte on the measurement of the [analyte] quantity" (McNaught and Wilkinson, 1997). As the sample consists of a complex mixture including potentially concentrated compounds, matrix effect can result in decreased ionization of some analytes (ion-suppression). Although the exact mechanism of matrix effect is unknown, competition for available charges at the surface of the droplets has been suggested (Cech and Enke, 2000). Recently, Nanita (2013) proposed an exponential model for the matrix effects, $ME_A$, which depends on $P(t)$. The model of the flowgram for analyte $A$ then becomes: $P_A(t) = k_A P(t) - ME_A(P(t))$ (Fig. 1).

Finally, the signal from the analyte can be contaminated by compounds from the carrier solvent. As the carrier flow is constant throughout the experiment, the contribution of solvent compounds to the intensity of analyte $A$ can be modeled by a constant baseline, $B_A$. The final equation for the flowgram intensity of analyte $A$ is:

$$I_A(t) = k_A \cdot P(t) - ME_A(P(t)) + B_A + \epsilon \quad (1)$$

where $\epsilon$ is the electronic (heteroscedastic) noise from the instrument (Fig. 1; notations are summarized in Table 1).

**Table 1.**
Notations

| Symbol | Signification |
|---|---|
| $I_A$ | Observed intensity (of analyte $A$) |
| $k_A$ | Analyte response factor |
| $P$ | Normalized sample peak: it corresponds to the theoretical peak in the absence of matrix effect and baseline; it is identical for all analytes and depends on the injection conditions only |
| $ME_A$ | Matrix effect function (of analyte $A$) |
| $B_A$ | Baseline (of analyte $A$) |
| $l_A, u_A$ | Lower and upper time limits (of analyte $A$) |
| $\epsilon$ | (Heteroscedastic) noise |
| $V(I)$ | Noise variance model as a function of intensity |
| $\hat{E}_A$ | Sum of observed noise, $I_A - (k_A P - ME_A(P) + B_A)$, within $(l_A, u_A)$ |

By using additional acquisitions of each analyte $A$ individually, Nanita (2013) managed to quantify the $k_A \cdot P(t)$ terms in a biological matrix (i.e., the true analyte peak in the absence of matrix effect). In untargeted metabolomics, however, such a methodology cannot be used, and the nonconvex model described in Equation 1 cannot be solved for each individual flowgram due the high level of heteroscedastic noise at low intensity. An alternative strategy for quantification then consists in summing the intensities. As an example, Fuhrer *et al.* (2010) proposed to sum all intensities throughout the flowgram. This method, however, cannot discriminate between the analyte and the solvent signals ($B_A(t)$ term) in the absence of additional blank control acquisitions (which are not always available). In addition, detection of the peak time borders specific to each analyte signal is critical for accurate quantification.

Here, we therefore propose a new methodology for analyte quantification. The main idea is to estimate the sample peak by applying Equation 1 to a subset of flowgrams with good quality (i.e. high intensities, no baseline, and minimum matrix effect), and then to use this sample peak to perform matched filtration on each individual flowgrams. The full approach consists of four steps. First, the variance of the noise $\epsilon$ is estimated by using the whole acquisition (*Noise estimation*). Second, within each analyte flowgram, the solvent baseline $B_A(t)$ is estimated. Third, the lower $l_A$ and upper $u_A$ time limits of the analyte signal are determined: the normalized sample profile, $P(t)$, is approximated by regressing the $kP(t) - ME(P(t))$ model on a subset of the flowgrams with high maximum intensity, low matrix effect and no baseline (*Sample peak modeling*). $P(t)$ is then used to find $l_A$ and $u_A$ in each flowgram by matched filtering (*Integration*). In cases of strong matrix effect, the central part of the analyte signal may flatten or even curve downward: in such situations, $l_A$ and $u_A$ are refined by looking for a second intensity maximum (with either $P(t)$ or a triangular wavelet). The analyte intensities between $l_A$ and $u_A$ are then used to compute an indicator of matrix effect (correlation with $P$), and to perform integration. Fourth, a statistical test is performed to assess if the measured signal is significantly higher than the solvent baseline by using the estimated noise model (*Solvent Filtering*). In conclusion, the methodology has two advantages: 1) the fit is performed only on data which are not subject to strong matrix effect, and 2) the filter procedure provides efficient and robust detection of all individual peak borders.

Finally, after analyte detection and quantification in each raw file, m/z values are aligned between samples (*Peak grouping*) by using a kernel density estimation. At this stage, the feature by sample table of peak intensities is generated. Missing values may be imputed by either the *k*-Nearest-Neighbors (Hrydziuszko and Viant, 2012; Shah *et al.*, 2017) or Random Forest (Stekhoven and Bühlmann, 2012; Di Guida *et al.*, 2016) methods (*Imputation*). The peak table can be exported to different formats for downstream statistical analysis and annotation (*Software implementation*).

# Methods

Our approach for FIA-HRMS preprocessing was implemented as a suite of algorithms named *proFIA*, and further packaged in R (R Core Team, 2017). The main steps of the workflow are (Fig. 2):
1. Peak detection and quantification (within each sample file):
    a. *Band detection*: Detection of m/z bands containing analyte signal
    b. *Noise estimation*: Estimation of the variance of the noise (Wentzell and Tarasuk, 2014)
    c. *Estimation of the sample peak profile*: An exponentially modified Gaussian (Kolev, 1994; Nanita, 2013) is fitted to the bands with high intensities and low matrix effect and baseline
    d. *Quantification and solvent filtering*: For each m/z band, the peak borders in the time dimension are determined by using the sample peak model; The

intensity is computed as the area under the smoothed signal; The noise model is used to filter out features which are not signficantly distinct from noise
2. *Feature grouping* (between sample files) by using a kernel density in the m/z dimension
3. *Imputation* of missing values (within the peak table), either by a modified version of k-Nearest Neighbors which takes into account the truncation of data at the limit of detection (Hrydziuszko and Viant, 2012; Shah *et al.*, 2017), or by a Random Forest approach (Stekhoven, D.J. and Büuehlmann, 2012, Guida *et al.*, 2016)

The workflow takes as input centroided FIA-HRMS raw files in open format (mzML, mzXML, mzData, or NetCDF), and produces a single feature by sample table of analyte intensities for downstream statistical analysis, either in a format similar to the peak table from XCMS, or as an ExpressionSet R object, or in the '3 table' data and metadata format from the Workflow4Metabolomics Galaxy online infrastructure (Giacomoni *et al.*, 2015; http://workflow4metabolomics.org).

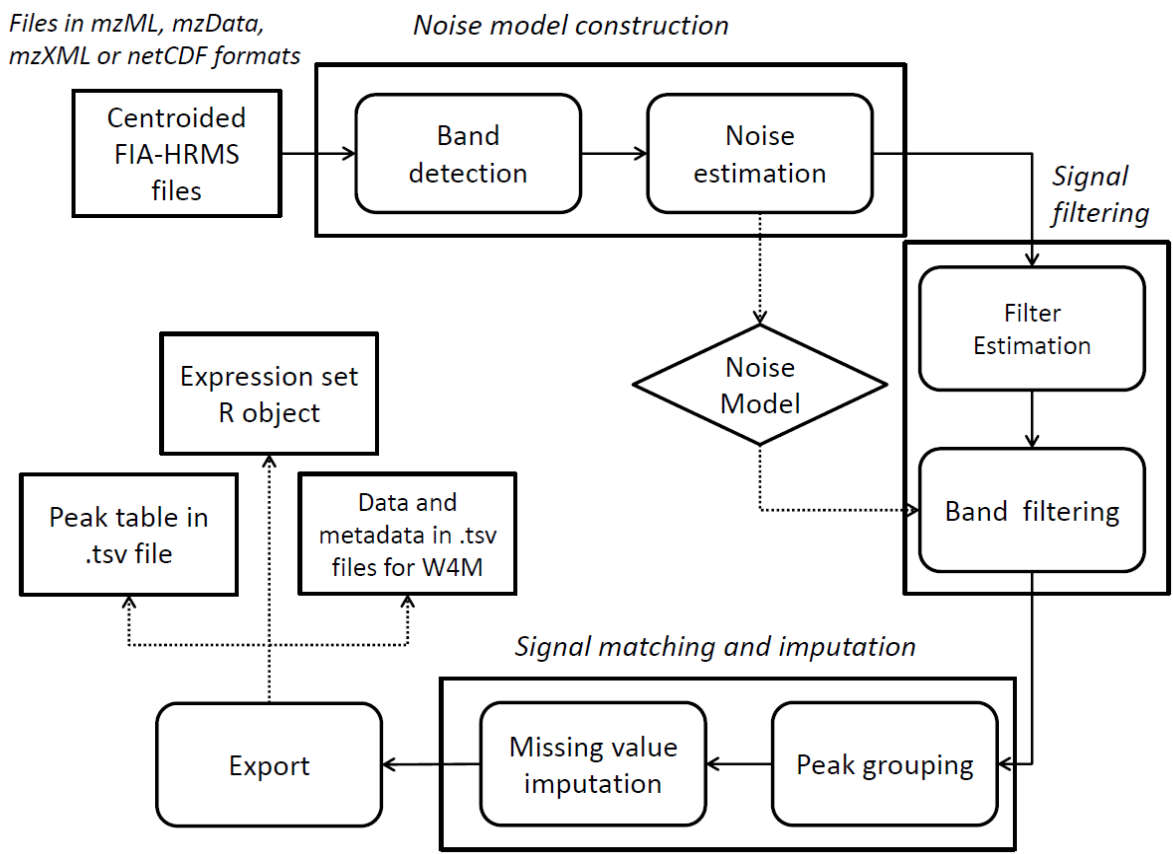

**Fig. 2. *proFIA* workflow.** Input files are centroided raw data in standard formats. First, within each sample file, mass bands are detected in the m/z by time plane (using a *ppm* parameter), and a noise model is built. The sample peak is modeled, and subsequently used within each m/z band to detect the time limits of the analyte signal, and evaluate its quality. Second, the previously detected features are grouped between samples by using a kernel density in the m/z dimension. Finally, a weighted k-Nearest Neighbors approach is implemented for the imputation of missing values. The output is the peak table containing the characteristics of each feature (intensity, mz limits, mean correlation with the sample peak).

## Estimation of the injection time window

Estimation of the time limits of the sample injection is required in the following step (*Band detection*) to discard bands with too little signal related to the injection (i.e. mainly consisting of noise). The injection window is estimated by applying geometric criteria to the smoothed Total Ion Flowgram (TIF; sum of all flowgrams). The details of the algorithm are described in the supplementary file S1.

## Band detection

Within each raw file, the detection of regions containing centroid signal has been used successfully in LC-HRMS to identify the m/z values from ions and discard electronic noise (Tautenhahn *et al.*, 2008). Preprocessing of FIA centroids must however address two additional constraints: 1) due to the matrix effect, some centroids may be missing within consecutive scans in an m/z trace, and 2) the density of centroids within each scan is high (no chromatographic separation) and may result in unresolved peaks. Our algorithm for the detection of m/z bands processes the scans one after another, and, for each centroid *c* (with m/z value *mz* and intensity *int*), either groups *c* with the closest band, or creates a new band (if there is no existing band within a *ppm* tolerance). The distance between a band *B* and *c* is defined by:

$$Dist(B, c) = \frac{|b_{mz} - c_{mz}|}{ppm \times c_{mz}} + \frac{|log(b_{int}) - log(c_{int})|}{n}$$

where *b* is the *B* centroid from the previous scan. The second (intensity) term aims at preventing the incorporation of electronic noise or centroids from isobars in band *B* (*n* is a parameter which is by default set to 2). The m/z bands cover the whole acquisition time range (to overcome constraint 1 described previously; Fig. 3). After detection, only bands with more than a *bandCoverage* proportion of centroids within the injection time window (default is 30%), or with more than *sizeMin* consecutive centroids (default set to half the size of the injection time window), are selected.

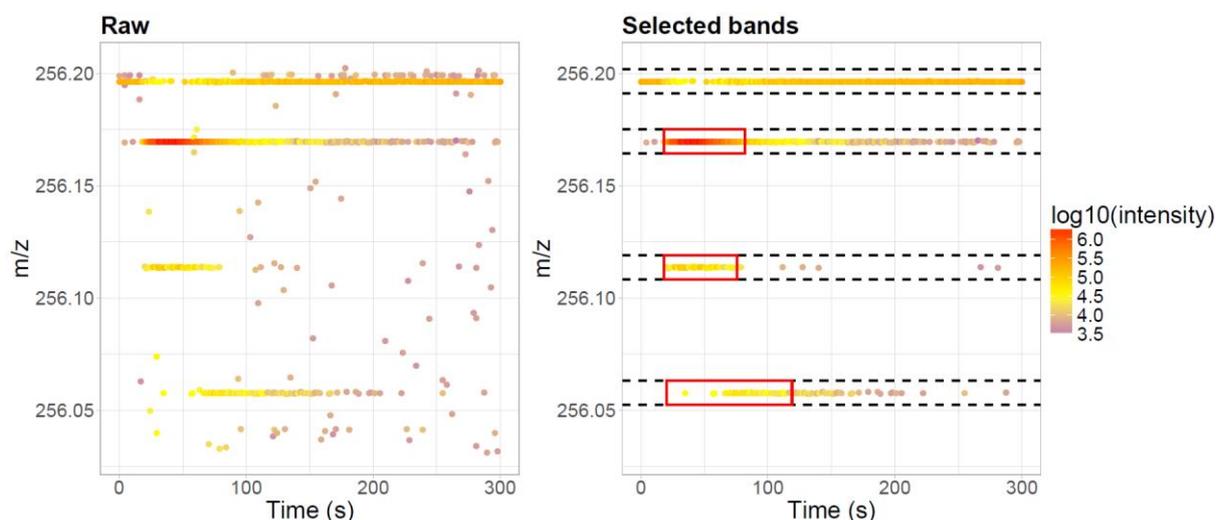

**Fig. 3. Detection of m/z bands containing analyte signal.** (**A**) Raw data within the 256.05-256.20 m/z region. (**B**) The bands detected by the *proFIA* algorithm are shown with black dots (the m/z width

of the bands is set to 0.01 for visualization purpose). During the following steps (*Integration and Solvent Filtering* section), the time limits of each analyte signal are detected (red boxes); when the signal is not significantly distinct from the solvent, the band is discarded (such as the band close to m/z = 256.20 in this example).

## Noise estimation

Determining the noise variance as a function of intensity (denoted $V$), is essential in the two subsequent steps of the workflow (i.e. for the estimation of the sample peak by least-square regression, and for the statistical testing of the difference between the putative analyte signal and the solvent baseline). To build the $V$ model, signals from the previously detected bands are used. The noise variance was first estimated at consecutive levels (bins) of intensity, as the mean of the squared difference between the flowgram signal and its smoothed profile obtained with a Blackman Windowed sinc Filter (Wentzell and Tarasuk, 2014). The variance $V(I)$ is then modeled as the loess regression of the noise variance on the intensity (see the model obtained with the metabolomics acquisition in the supplementary file S1).

## Sample peak modeling

A model of the sample peak $P$ is required for the accurate determination of analyte time limits in each flowgram (next section). Since Equation 1 cannot be solved for each individual flowgram (because there is usually not enough data points compared to the levels of noise and baseline), we implemented a two-step strategy relying on the flowgrams of highest quality to estimate the sample peak.

First, flowgrams with no solvent, minimal shift and minimal matrix effect (i.e. maximum intensity) are selected. We denote this set of flowgrams $M$. Second, the following model is fitted to these flowgrams:

$$I_A[i] = k_A \cdot P_{\mu,\sigma,\tau}[i] - ME_{a_A,b_A}(P_{\mu,\sigma,\tau}[i]) + \epsilon_I[i]$$

where
- $P_{\mu,\sigma,\tau} = G_{\mu,\sigma} * E_\tau$ is an exponentially modified Gaussian (Kolev, 1994); parameters $\mu$ and $\sigma$ are the mean and standard deviation of the Gaussian, and $\tau$ is the decay of the Exponential; these parameters are identical for all flowgrams,
- $ME_{a_A,b_A}$ is a first order exponential model of matrix effect (Nanita, 2013); the parameters $a_A$ and $b_A$ are specific to each flowgram,
- $\epsilon_{I[i]} \sim N(0, V(I[i]))$ is the noise model estimated previously.

The vector of parameters $\theta_A = (\mu, \sigma, \tau, a_A, b_A)$ is estimated by least squares fitting (comparison with Maximum Likelihood estimation, which allows to take into account noise heteroscedasticity, gave similar results, since only high intensity signals are used, resulting in low levels of noise compared to signal).

Regression on all $M$ flowgrams is performed simultaneously in order to obtain a robust estimation of the sample peak. The minimization of:

$$\sum_{j \in M} |I_j - k_j \cdot P_{\mu,\sigma,\tau} + ME_{a_j,b_j}(P_{\mu,\sigma,\tau})|$$

was achieved with the Levenberg-Marquardt algorithm (Moré, 1978) implemented in the *minpack.lm* package (Elzhov et al., 2016). Starting values are determined by using the TIF peak for parameters $\mu, \sigma, \tau$, and are set to 0.05 and 5 for $a$ and $b$, respectively.

## Integration and solvent filtering

The last step of peak detection and quantification consists in determining within each flowgram the time borders of the signal (step 1), in quantifying the analyte by integrating the flowgram within this window, and, in cases where a solvent baseline is present, in assessing whether the quantified signal is significantly different from the solvent (step 2). As these steps have not been described for FIA signals yet, we designed the following algorithms (Fig. 4), which are used on each detected band to facilitate reading, the $A$ subscript has been omitted from the variable names $l, u, I, B, \epsilon$ and $E$ in this section).

Step 1: Time values corresponding to the signal borders $(l, u)$ and apex are obtained by convolution of the flowgram profile with the sample peak model $P$ (Fig. 4; red curve). To refine the lower limit $l_A$ in case of strong signal suppression due to matrix effect (about 5% of the flowgrams in our metabolomics data set; see the *Results* section), an additional filter based on a triangular wavelet is used (Fig. 4B; yellow line).

Step 2: In case where solvent is present, a statistical test is used to discard the analyte if the signal is not significantly different from the baseline. The null hypothesis $H_0$ states that the signal consists of baseline only: $I = B + \epsilon$, where the noise $\epsilon$ is modeled at each time $i$ by $\epsilon_i \sim N(0, V(B[i]))$ with the $V$ function estimated previously. Under the assumption of noise independence with time, the total noise between the signal limits becomes:

$$E = \sum_{i=l}^{u} \epsilon_i \sim N(0, \sum_{i=l}^{u} V(B[i]))$$

The $\hat{E} = \sum_{i=l}^{u}(I[i] - \hat{B}[i])$ statistic (with $B$ being the linear segment between the peak limits) is therefore compared to the $N(0, \sum_{i=l}^{u} V(B[i]))$ distribution. The feature intensity is set to $\hat{E}$ if the unilateral test *p*-value is inferior to a given threshold (0.01 by default), or set to NA otherwise.

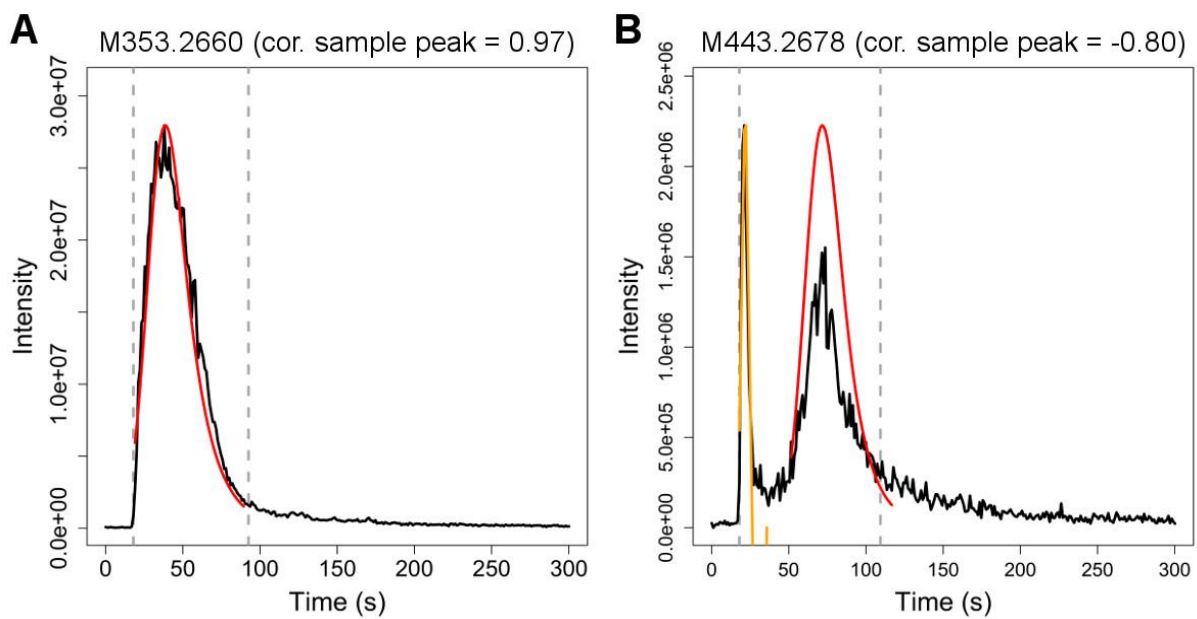

**Fig. 4. Determination of the time limits of the analyte signal.** (**A**) The lower and upper time limits are computed by matched filtering of the sample peak model (red). (**B**) In case of signal suppression, the lower limit is refined by using a triangular wavelet (yellow).

Two additional metrics are computed to assess the quality of the analyte signal. When the time limits are shifted to higher values compared with the sample peak in at least one of the samples (as the result of retention in the injection pipe), the feature is flagged. In the absence of shift, the mean correlation between the analyte signal in each sample and the sample peak is computed: a low correlation (i.e. < 0.2) tends to indicate that the feature signal is strongly affected by matrix effect in some of the samples. Furthermore, a diagnostic plot describing the quality of the detected features is provided (see the *Software implementation* section and the supplementary file S1).

## Peak grouping

After feature detection and quantification within each file, m/z values are aligned between the samples. We use a kernel density estimation to cluster the data of similar mass-to-charge. Compared with the *xcms.group* method (Smith et al., 2006), the density kernel is applied here in the m/z dimension. This type of approach is well suited to Gaussian measurement errors, and requires only one parameter (bandwidth of the density function). At the end of the grouping, the feature by sample table of peak intensities is built.

## Imputation

Some of the missing values in the peak table may result from a technical failure to detect a compound in a sample (e.g. for concentrations close to the limit of detection). An imputation step is therefore useful in the preprocessing workflow. The imputation approach based on integration of raw signal within the expected region of interest (as implemented in the *xcms.fillPeaks* algorithm; Smith et al., 2006) may fail to detect any signal in case of high-resolution data, because the detected bands are thin. As an alternative, a *k*-Nearest Neighbor method (k-NN), which imputes a missing value as the average of the closest features from the peak table, was shown to be optimal for DI-MS data (Hrydziuszko and Viant, 2012). Such an approach requires that features be scaled before nearest neighbors are computed. To refine the estimation of the intensity distribution for feature scaling, Shah et al. (2017) recently proposed to take into account the limit of detection of the instrument by using a truncated normal distribution model (KNN-TN). The described methodology, however, assumes that the distribution can be modeled by a single Gaussian, which may lead to errors in case of multiple sample classes. We therefore implemented a modified KNN-TN imputation method where the similarity with neighbors is computed for the samples of the same class only. As Random Forest was recently shown to achieve efficient imputation of missing values in MS data sets (Gromski et al., 2014; Di Guida et al., 2016), this alternative imputation method (Stekhoven and Bühlmann, 2012) was also included in *proFIA*.

## Software implementation

The complete workflow was implemented in R (R Core Team, 2017), and is available as the *proFIA* package (http://bioconductor.org/packages/proFIA) from the Bioconductor repository (Gentleman *et al.*, 2004). The successive functions are described in the supplementary file S1 (see the package vignette for a practical example of an application to a real data set). *proFIA* is also available as a Galaxy module on the Galaxy Tool Shed

(https://toolshed.g2.bx.psu.edu/), and on the Workflow4Metabolomics online platform (http://workflow4metabolomics.org) which provides advanced and user-friendly features for workflow management (Giacomoni *et al.*, 2015). A subset from the FIA-HRMS Orbitrap Fusion data described in this study (6 files out of 18) is available as the *plasFIA* package (http://bioconductor.org/packages/plasFIA).

# Results

We evaluated the quality of the preprocessing by the *proFIA* workflow on a real FIA-HRMS dataset consisting of human serum either pure or spiked with a pool of 40 compounds belonging to a large variety of chemical classes (supplementary file S2) at 5 increasing concentrations (from C1 = 10 ng/mL to C100 = 1 µg/mL; C0 corresponds to the unspiked serum sample). The samples were analyzed in triplicate with an Orbitrap *Fusion* mass spectrometer at a mass resolution of 500K (Thermo Fisher Scientific; supplementary file S1). The resulting 18 raw files were converted into the open mzML format and centroided with the proteoWizard software (Chambers et al., 2012). Preprocessing of the raw files with the *proFIA* package (*ppm* = 2, *dmz* = 0.0005, and *fracGroup* = 3/18) resulted in a peak table of 1,082 features (supplementary file S3).

**Table 2.**
Reproducibility of peak quantification

| Intensity range | 1st Quintile F: [1.0 x $10^4$, 9.2 x $10^4$] E: [9.9 x $10^2$, 6.0 x $10^3$] | 2nd Quintile F: [9.2 x $10^4$, 1.4 x $10^5$] E: [6.0 x $10^3$, 9.8 x $10^3$] | 3rd Quintile F: [1.4 x $10^5$, 2.1 x $10^5$] E: [9.8 x $10^3$, 1.5 x $10^4$] | 4th Quintile F: [2.1 x $10^5$, 4.2 x $10^5$] E: [1.5 x $10^4$, 3.0 x $10^4$] | 5th Quintile F: [4.2 x $10^5$, 7.8 x $10^7$] E: [3.0 x $10^4$, 2.3 x $10^7$] |
|---|---|---|---|---|---|
| $CV_F$ (%) | 16.1 | 13.1 | 11.6 | 10.9 | 7.7 |
| $CV_E$ (%) | 10.0 | 9.1 | 8.7 | 9.2 | 9.1 |

The coefficient of variation (CV) between the replicates is computed within consecutive intensity windows (quintiles) along the dynamic range for the *Fusion* (F) and *Exactive* (E) Orbitrap acquisitions.

The reproducibility was first assessed by comparing the peak detection and quantification between replicates. Sixty-seven percent of the features were detected in all 3 triplicates (supplementary file S1). Peak quantification was also reproducible, with a mean coefficient of variation (CV) for all the intensities along the dynamic range of the instrument of 12% (Table 2). Furthermore, the high reproducibility of *proFIA* was shown to outperform the *centWave* algorithm (Tautenhahn *et al.*, 2008; supplementary file S1).

To validate that *proFIA* can be applied to data from distinct analytical platforms, a second dataset was used, which consisted in the analysis (in triplicate) of 8 dilutions of a serum sample on an Orbitrap *Exactive* instrument with a resolution set to 100K (Thermo Fisher Scientific, supplementary file S1). The preprocessing of the resulting 24 raw files by *proFIA* (*ppm* = 8, *dmz* = 0.001, *fracGroup* = 3/24) generated a peak table of 2,046 features. The reproducibility of peak quantification proved similar to the results from the Orbitrap *Fusion* (mean CV of 9%; Table 2). To further evaluate the preprocessing on a distinct biological matrix, *proFIA* was applied to urine samples (quality control pools) from a recent cohort study analyzed by FIA coupled to an LTQ Orbitrap XL with a resolving power set to 60K (Habchi et al., 2017). Under these conditions, 62 out of the 66 features selected by chemometrics analyses in the original study (i.e., 94%) were detected by *proFIA* in both ionization modes (the four missing features having very low intensities in these pools), thus demonstrating the value of the workflow for multiple biological matrices.

**Table 3.**
Confusion matrix of the automated peak detection.

|  |  | *proFIA* | | |
|---|---|---|---|---|
|  |  | Yes | No | Total |
| Manual | Yes | 460 | 6 | 466 |
|  | No | 15 | 239 | 254 |
|  | Total | 475 | 245 | 720 |

Automated preprocessing with *proFIA* was then compared to manual integration by using the vendor software (Xcalibur, Thermo Fisher Scientific). High precision (96%; proportion of peaks detected by *proFIA* which were also detected manually) and recall (98%; percent of peaks detected manually which were also detected by *proFIA*) values were obtained for all spiked compounds from the *Fusion* dataset (Fig. 5 and Table 3). It should be noted that the majority of the 15 peaks detected only by *proFIA* (false positives) had intensities close to the limit of detection (see the supplementary file S4 for the corresponding flowgrams). Peak quantification by *proFIA* was also very precise, with a mean intensity difference of less than 5% compared to manual integration.

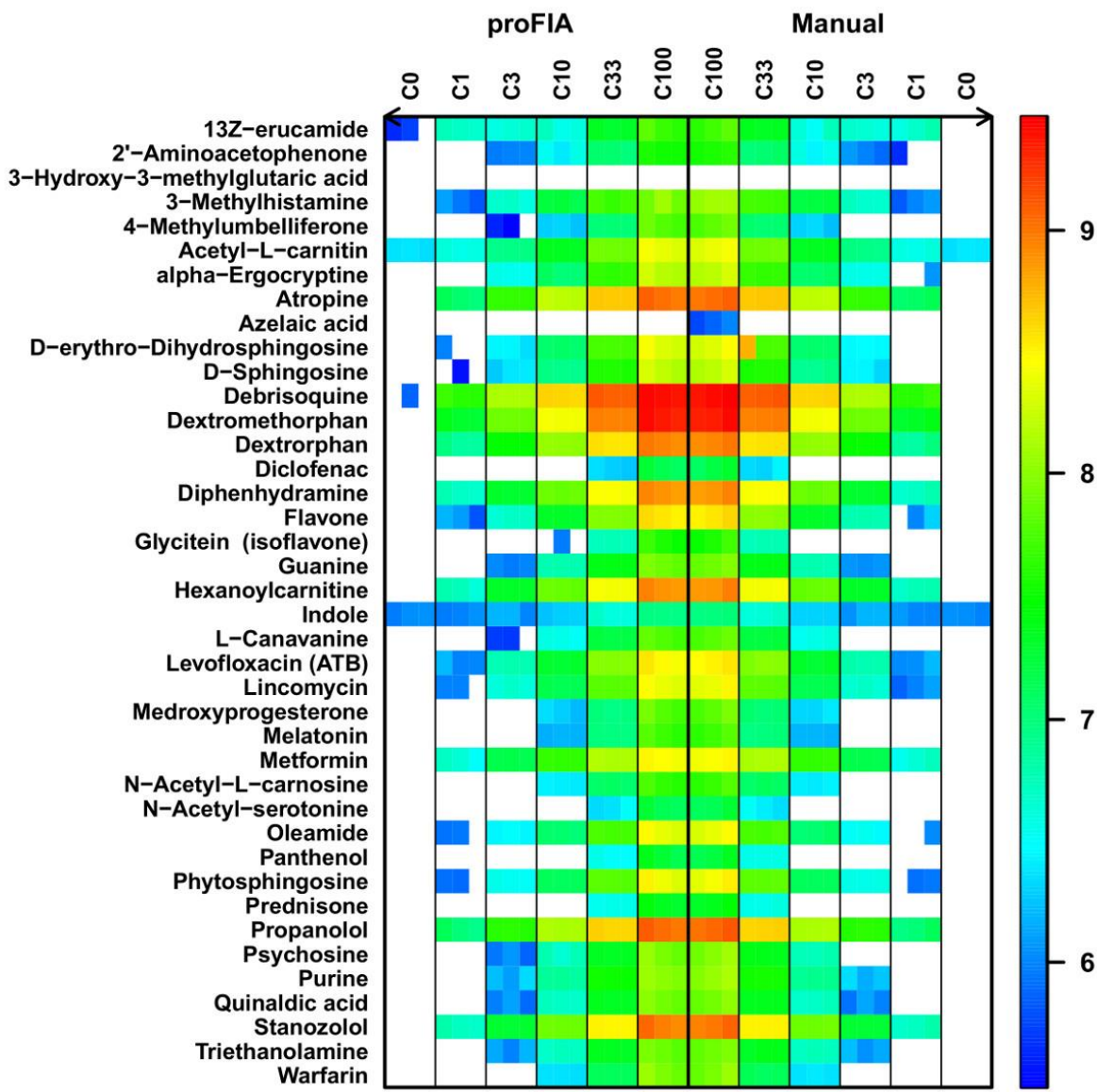

**Fig. 5. Evaluation of peak detection and quantification.** For each of the 40 compounds spiked in the serum samples at various dilutions and analyzed in triplicate (*Fusion* dataset; see the supplementary files S1), the automated quantification with *proFIA* (left) was compared to the manual integration of peaks with the vendor software (right). The concentration of the spiking mixture is indicated in the sample label (C1 = 10 ng/mL). The white color denotes the absence of signal.

Finally, we performed statistical analysis and annotation of the peak table generated by *proFIA* to illustrate the value of a comprehensive FIA-HRMS analysis workflow. First, multivariate modeling of the spiking concentration was performed by using Partial Least Squares (Thévenot et al., 2015) and confirmed that spiked compounds had high loading values (supplementary file S1). Second, annotation of the features was performed by computing m/z differences (for isotopes and adducts), followed by matching to the Human Metabolome Database (Wishart et al., 2007; mass tolerance set to 0.001; Table 4 and supplementary file S6). A total of 300 (28%) features matched HMDB metabolites, including 211 (19%) matching a single chemical formula. As expected, similar annotation performances were obtained with the *Exactive* platform (Table 4). Altogether, these results highlight the importance of the FIA-HRMS approach and the *proFIA* preprocessing for untargeted metabolomics.

**Table 4.**
Putative annotations of the m/z values detected in serum samples.

| Mass Spectrometer | Single match on HMDB | Multiple matches on HMDB | Supplementary Adducts | Isotopes | Features not found | Total number of features |
|---|---|---|---|---|---|---|
| Fusion | 211 (20%) | 89 | 15 | 137 | 645 | 1,082 |
| Exactive | 303 (15%) | 105 | 35 | 192 | 1,411 | 2,046 |

# Discussion

To preprocess data from Flow Injection Analysis coupled to High-Resolution Mass Spectrometry (FIA-HRMS), we developed a suite of algorithms which address the specificities of FIA-HRMS signal. In particular, the detection of m/z bands copes with the high-density of centroids in the m/z dimension. Furthermore, the time limits of each analyte signal are precisely estimated by relying on a model of the common sample peak and taking into account matrix effect. Finally, analyte signals are discriminated from solvent based on a statistical test comparing the difference between the intensity and the baseline to a model of the heteroscedastic noise. The full workflow was implemented in R and is available as the *proFIA* package on the bioconductor repository: it is fast (approximately 15 s per file), and includes parallel computation functionalities (Morgan, 2017). The value of the main parameters (*ppm* and *dmz*) can be easily inferred from the mass resolution of the instrument (see the package vignette). Furthermore, several metrics and diagnostic plots are provided to readily assess the quality of the overall experiment and of the individual features (including potential alteration of peak shape due to matrix effect, retention in the injection pipe, or presence of a solvent baseline).

High reproducibility of peak detection and quantification by *proFIA* was demonstrated with metabolomics datasets consisting of human serum analyzed on two distinct platforms (*Fusion* and *Exactive* Orbitrap instruments). Importantly, the automated *proFIA* preprocessing achieved performances similar to the time-consuming manual integration of peaks with the vendor software. The *proFIA* workflow significantly improves our previous results with an LTQ-Orbitrap (Madalinski *et al.*, 2008) both in terms of the number of detected compounds (1082 vs 400) and of the coefficient of variation between replicates (below 21% vs more than 50%). Moreover, our approach now includes a discrimination between analyte signal and solvent. Finally, the *proFIA* workflow proved relevant when applied to urine samples, thus highlighting the value of the approach for various kinds of key biological matrices.

A quality metric based on the correlation between the analyte signal and the sample peak was implemented to detect features whose signal shape (and hence intensity) are strongly affected by matrix effect. Recently, Nanita (2013) described a methodology to estimate the true height of each analyte peak even in cases of signal suppression. The approach is based on the hypothesis that at the borders of the sample peak time window, the amount of ions entering the mass spectrometer is reduced, resulting in a matrix effect close to 0 (infinite dilution). Future analytical developments to further reduce the noise level and increase the number of scans may allow such an approach to be implemented in non-targeted metabolomics.

Recently, FIA-MS was applied to high-throughput metabolomics profiling in bacteria, with up to thousand acquisitions per day (Fuhrer *et al.*, 2011; Sevin *et al.*, 2017). In this study, a lower resolution was used (10,000). In fact, the authors relied on the extensive characterization of *E. coli* in metabolite databases such as KEGG to characterize the metabolome variations. In human samples, however, discrimination between isobaric compounds (and within a list of putative adducts) by using high-resolution is critical to achieve high-throughput annotation in large cohort studies. Here, 300 molecules were matched to the HMDB database, including 211 (19%) with a single formula. Approaches combining FIA-MS and MS/MS experiments should further improve the structural characterization of the detected metabolites.

In conclusion, we have developed the first open-source software for the preprocessing of FIA-HRMS data. Availability of data and source code is critical to improve computational workflow reproducibility and comprehensiveness, and to benchmark new algorithms. The *proFIA* package, which has also been integrated into the user-friendly Workflow4Metabolomics online infrastructure (Giacomoni *et al.*, 2015), thus opens new opportunities for high-throughput phenotyping.

# Acknowledgements

We thank Sandra Alves and Estelle Rathahao-Paris for sharing the urine data, and Jean Claude Tabet for his reviewing of the manuscript. This work has been supported by the Agence Nationale de la Recherche (MetaboHUB national infrastructure for metabolomics and fluxomics, ANR-11-INBS-0010 grant).

# Supplementary data

# *proFIA*: A data preprocessing workflow for Flow Injection Analysis coupled to High-Resolution Mass Spectrometry

Alexis Delabrière[1], Ulli M. Hohenester[2], Benoit Colsch[2], Christophe Junot[2], François Fenaille[2] and Etienne A. Thévenot[1]

[1]CEA, LIST, Laboratory for data analysis and systems' intelligence, MetaboHUB, F-91191 Gif-sur-Yvette, France
[2]CEA, DRF/JOLIOT/SPI, Laboratoire d'Etude du Métabolisme des Médicaments, MetaboHUB, F-91191 Gif-sur-Yvette, France

# Supplementary methods

## Estimation of the injection time window

Estimation of the time window corresponding to the injection of the sample is performed as follows:
1. The Total Ion Flowgram (TIF) is normalized both in the time and intensity dimensions
2. The solvent value is set to the first intensity of the TIF
3. The starting time of the injection (lower limit of the injection window: $l_{inj}$) is computed as the first time corresponding to 3 successive intensities superior to the solvent
4. The time of maximum intensity of the smoothed TIF (median filter) is set to $m_{inj}$.
5. All time points superior to $l_{inj} + m_{inj}$ are considered as candidates for the end of injection (upper limit: $u_{inj}$). For each candidate time $u$, let us denote $M$, $U$, and $L$ the summits of the triangle joining the signal intensities at time points $m_{inj}$, $u$ and at the last time point of the flowgram. $u_{inj}$ is computed as the time point maximizing $cos(\overline{UM}, \overline{UL}) - UM.UL$ (the second term avoids the selection of points with noisy intensities at the end of the flowgram)

## Noise model

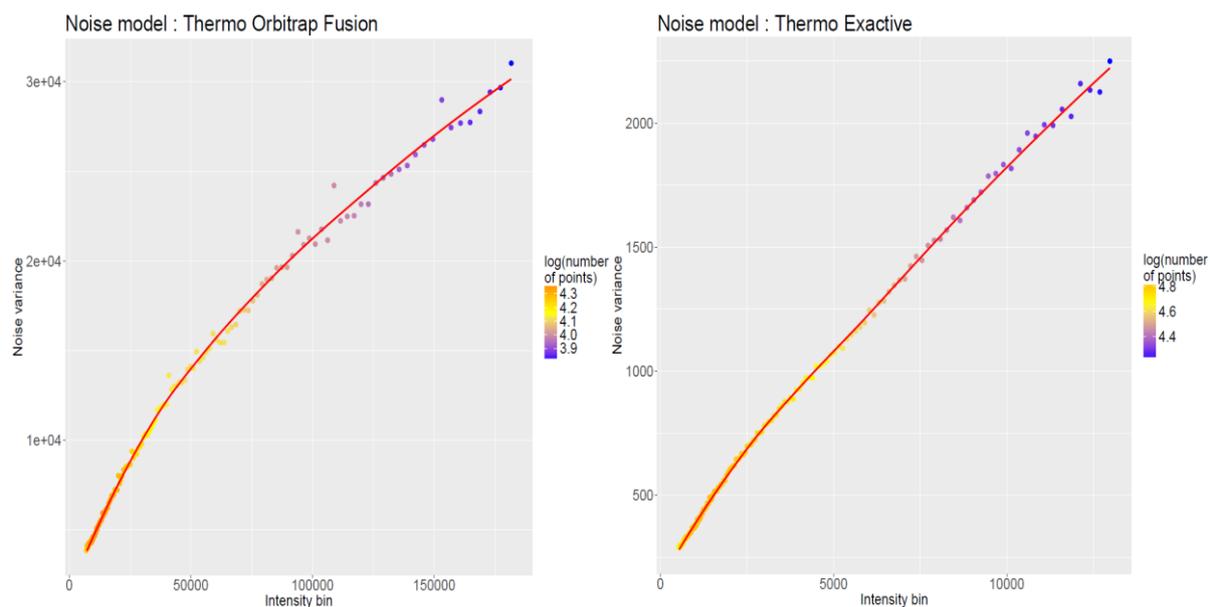

**The model of the noise variance as a function of intensity estimated on the datasets obtained with the Orbitrap *Fusion* (left) and *Exactive* (right).**

# Solvent filtering: validity of the statistical test even in cases where the signal is affected by matrix effect

Here we demonstrate that the *p*-value used to discriminate signal from solvent, which was obtained by modeling the baseline as a linear segment, can still be used when a matrix effect is taken into account. Our hypothesis $H_0$ still states that the signal consists of baseline only. Since we now include matrix effect, our equation describing intensity under $H_0$ now becomes:

$$I = B - ME(P) + \epsilon$$

Our previous linear estimation of $B$ is superior to the quantity $B - ME(P)$, as the contribution of matrix effect is always negative. As a result, our previous estimate of the noise at each data point $i$, $\hat{\epsilon}_i = I - \hat{B}$ is inferior or equal to the real value of $\epsilon_i = I - B$, and our previous test statistic $\hat{E} = \sum_i \hat{\epsilon}_i \leq E$ is always under-estimated (red cross on the figure below). The calculated *p*-value is therefore biased upward:

$$P_{H_0}(X > E) \leq P_{H_0}(X > \hat{E})$$

Moreover, since the variance estimation $V$ is an increasing function, $\hat{B} > B$ results in $V(\hat{B}) > V(B)$. Consequently, the variance of $E$ is over-estimated.

Finally, let us consider the following property of Gaussian distributions: if $X$ and $Y$ are random variables following two gaussian distributions of mean 0 and standard deviations $\sigma_X \geq \sigma_Y$, we have for all $U$ superior to 0 :

$$P(X > U) \geq P(Y > U)$$

Since the standard deviation of $\hat{E}$ is inferior to the true standard deviation of $E$, the above property results in $P_{H_0}(X > \hat{E}) \geq P_{H_0}(Y > \hat{E})$ with $X \sim N(0, \sum_i V(\hat{B}_i))$ and $Y \sim N(0, \sum_i V(B_i))$. The *p*-value calculated previously under the hypothesis $I = B + \epsilon$ (i.e. in the absence of matrix effect) is therefore a majorant of the *p*-value that would be obtained by considering the model affected by the matrix effect. In conclusion, using the *p*-value obtained without considering the matrix effect does not increase type II errors, i.e. errors resulting in a signal considered as an "analyte" whereas it is in fact the baseline solvent (to our experience there are few additional errors of type I, and such errors are not limiting for the quality of the preprocessing).

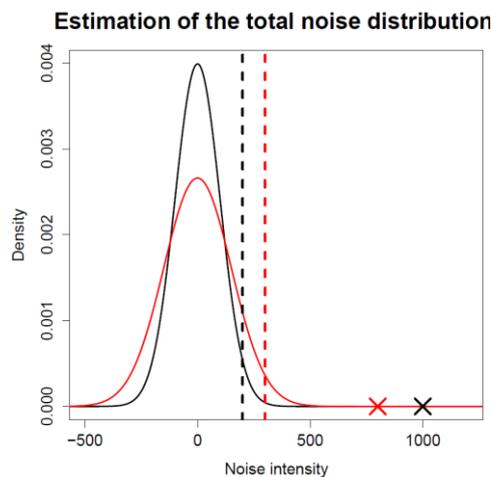

**Noise distributions and test statistics under $H_0$.** The distribution and test statistic (cross) estimated in the absence of matrix effect are shown in red. The true distribution and statistic are in black.

## Software implementation

The main functions, parameters, input and output formats from the *proFIA* workflow are (see the figure below):
1. Mass band detection within each raw files (centroid mode; mzML, mzXML, mzData, or NetCDF formats) with the *proFIAset()* function (the *ppm* and *dmz* parameters control the maximum m/z width of the bands; *ppm* is the relative value, and *dmz* is the absolute minimum value, since the accuracy of mass instruments decreases at low masses). This step includes noise model building (with an internal call to the *estimateNoiseMS()* method), sample peak modeling and band filtering (with *findFIASignals()*). The *plotRaw()* method can be used to visualize the raw data in the time by m/z plane. The peak models for all samples can be plotted with *plotSamplePeaks()*,
2. Grouping of the features between the samples in the m/z dimension with *group.FIA()*; the *ppmGroup* and *dmzGroup* parameters control the standard deviation of the kernel density in the m/z dimension; the *fracGroup* defines the minimum proportion of samples with non-zero intensity required in at least one class for a feature to be kept. The flowgrams corresponding to a given m/z value (and tolerance) can be plotted with *plotFlowgrams()*,
3. Building the peak table with *makeDataMatrix()*,
4. Optional imputation of missing values with *impute.FIA()*. Two methods are available: either modified K-Nearest Neighbors for truncated distributions (default), or Random Forest.

For convenience, all steps of the workflow have been wrapped into a single function, *analyzeAcquisitionFIA()*. The table with all feature characteristics (including m/z value and sample intensities) can then be exported with the *exportPeakTable()* method. The pre-processed data can also be exported as an ExpressionSet object (Bioconductor class for omics data sets; *exportExpressionSet()*), or in the '3 table' data and metadata tabular file format from the Workflow4Metabolomics Galaxy online infrastructure (*exportDataMatrix()*, *exportSampleMetadata()*, and *exportVariableMetadata()*).

Further details and examples are provided in the package vignette.

```
                    centroid files (mzML, mzXML, NetCDF, mzData)

analyzeAcquisitionFIA()     pset <- analyzeAcquisitionFIA(path, ppm = 2, dmz = 0.0005,
Wrapper                     ppmGroup = 1, dmzGroup = 0.0005, fracGroup = 0.2, k = 2)
```

**proFIAset()** — Peak picking
```
pset <- proFIAset(path, ppm = 2, dmz = 0.0005)
ppm:   maximum deviation between scans
       during band detection
dmz:   minimum threshold for the m/z deviation
       tolerance
```

```
plotRaw(pset, type = "raw"/"peaks", sample = 3, ylim =
plasMols[7, "mass_M+H"] + c(-0.1, 0.1), size = 0.6)
```

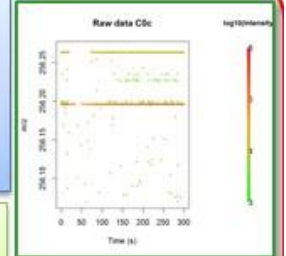

**group.FIA()** — Alignment    **plotSamplePeaks**(pset)

```
pset <- group.FIA(pset, ppmGroup=1, dmzGroup=0.0005, fracGroup=0.2)
ppmGroup, dmzGroup:   should be ≤ to the corresponding value
                      in proFIAset
fracGroup:   minimum fraction of samples with detected peaks in at least
             one class for a group to be created
```

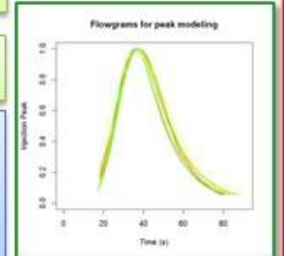

```
plotFlowgrams(pset, mz = plasMols[4, "mass_M+H"])

lMatch <- findMzGroup(pset, plasMols[, "mass_M+H"], tol = 3)
```

**makeDataMatrix()** — Formatting
```
pset <- makeDataMatrix(pset)
```

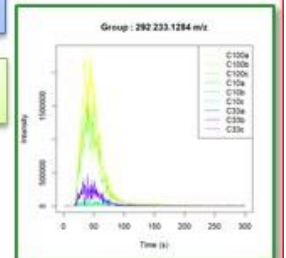

**impute.FIA()** — Imputation
```
pset <- impute.FIA(pset, k = 2)
k: number of neighbor features
```

**plot()** — Diagnostics    **plot** (pset)

**exportExpressionSet()** — Export
```
eset <- exportExpressionSet(pset)

peakTableDF <- exportPeakTable(pset)

dataMatrixMN <- exportDataMatrix(pset)

sampleMetadataDF <- exportSampleMetadata(pset)

variableMetadataDF <- exportVariableMetadata(pset)
```

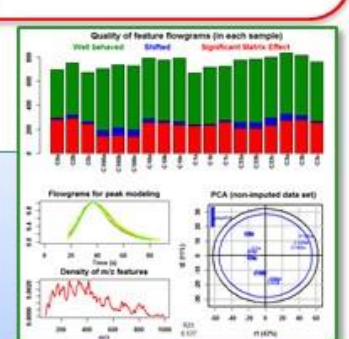

**Main methods, parameters, and outputs of the *proFIA* package.** Example of an application to the *plasFIA* dataset (serum samples analyzed with an Orbitrap *Fusion* at a mass resolution of 500K; the data are freely available as an Experiment package on Bioconductor). For the second dataset analyzed in the article (Orbitrap *Exactive* at a mass resolution of 100K), the *proFIA* parameters were: *ppm* = 8, *dmz* = 0.001, *ppmGroup* = 4, *dmzGroup* = 0.0005, as described in the Table S1 below. Please note that due to the particular design of these datasets (multiple dilutions injected in triplicates), the *fracGroup* value has been decreased to 0.2, and the imputation step is not relevant (and should be omitted).

## Metabolomics datasets

### Biological material and sample preparation

The serum sample was purchased from Biopredic (Rennes, France) and metabolites were recovered after methanol precipitation of proteins as described in Boudah et al. (2014)[1]. The spiking mixture (*plasFIA* dataset) contains 40 compounds which are described in the supplementary file S2. Concentrations of the serum sample and the spiking mixture are detailed in the Table S1 below.

### FIA-HRMS acquisitions

Samples were analyzed in triplicates. Details of the acquisition conditions are provided in Table S1.

---

[1] Boudah, S.; Olivier, M.-F.; Aros-Calt, S.; Oliveira, L.; Fenaille, F.; Tabet, J.-C.; Junot, C. Annotation of the human serum metabolome by coupling three liquid chromatography methods to high-resolution mass spectrometry. *Journal of Chromatography B,* **2014**, *966*, 34-47.

**Table S1: Sample preparation and acquisition conditions for the two metabolomics data sets.**

|  |  | *Fusion* dataset | *Exactive* dataset |
|---|---|---|---|
| Biological sample and spiked mixture | Serum dilution(s) | 1/50 | 1/3, 1/6, 1/12, 1/24, 1/48, 1/96, 1/192, and 1/384 |
|  | Spiked mixture | 0, 10, 33, 100, 333 and 1000 ng/mL | none |
| FIA-HRMS acquisition | Instrument | Orbitrap *Fusion* | Orbitrap *Exactive* |
|  | Spray voltage | 3.5 kV | 4 kV |
|  | Ionization mode | positive | |
|  | Capillary temperature (°C) | 320 | 325 |
|  | Sheath gas flow | 15 | |
|  | Auxiliary gas flow | 8 | 12 |
|  | m/z range | 85-1,000 | 95-1,000 |
|  | Mass resolution | 500K at m/z 200 | 100K at m/z 200 |
|  | AGC target | $10^5$ | $3 \times 10^6$ |
|  | Ion injection time (ms) | 100 | 250 |
|  | Runtime duration (min) | 5 | |
|  | Mobile phase composition | 75% isopropanol in water + 0.1 % formic acid | |
|  | Flow rate (µL/min) | 50 | |
| *proFIA* preprocessing | *ppm* | 2 | 8 |
|  | *dmz* | 0.0005 | 0.001 |
|  | *ppmGroup* | 1 | 4 |
|  | *dmzGroup* | 0.0005 | 0.0005 |

# Supplementary results

## Reproducibility of peak detection

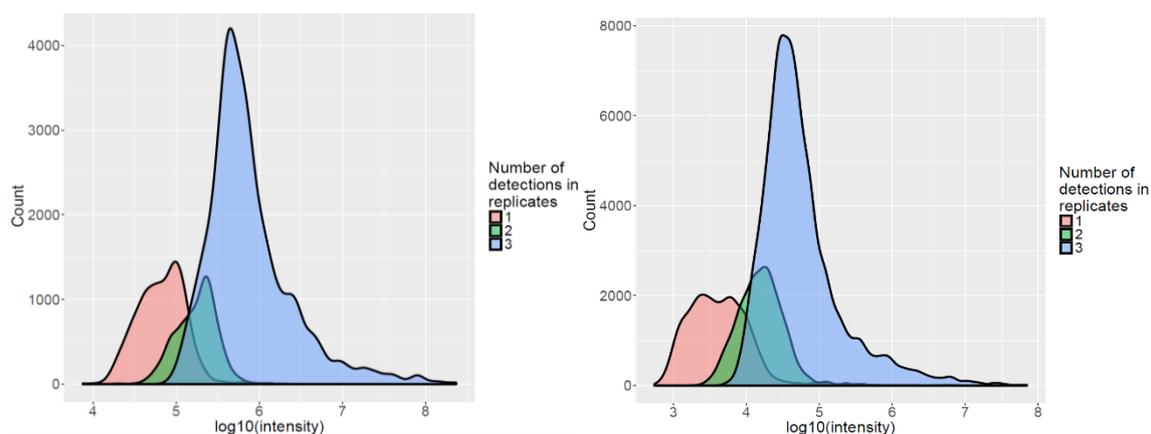

**Density of features simultaneously detected in the replicates analyzed with the Orbitrap *Fusion* (left) and *Exactive* (right).** Except for the features with the lowest intensities, a high reproducibility of peak detection is observed.

## Comparison with the *centWave* algorithm

The *centWave* algorithm[2] was compared to the *proFIA* results for the preprocessing of the FIA-HRMS *Fusion* dataset. The minimum value of the *peakwidth* parameter from *centWave* was set to a low value (i.e. 5) to take into account a potential matrix effect affecting the peak shape. The complete list of tested *centWave* parameter values is given below:

| Parameter | Values |
| --- | --- |
| *ppm* | 0.5, 1, 2, 3, 5 |
| *snthresh* | 3, 5, 10 |
| *prefilter (k)* | 2, 3, 4 |
| *prefilter (i)* | 100, 500, 1000 |
| *peakwidth (Max)* | 25, 50, 75, 100 |

Several parameters were also tested for the *proFIA* algorithm:

| Parameter | Values |
| --- | --- |
| *ppm* | 1, 2, 3, 4, 5 |

---

[2] Tautenhahn, R.; Bottcher, C.; Neumann, S. Highly sensitive feature detection for high resolution LC/MS. *BMC BioinformaticsI,* **2008**, *9* (1), 504.

| *bandCoverage* | 0.2, 0.3, 0.4, 0.5, 0.6, 0.8 |
|---|---|
| *pvalthresh* | 0.05, 0.01, 0.005, 0.001, 0.0001 |

To evaluate the performance of *proFIA*, we compared both the number of signals detected in the three replicates, and the mean coefficient of variation between the intensities. We selected features with intensities in the highest quintile because such features are expected to be the easiest to detect and accurately quantify by *centWave*. The results are summarized in the following figure:

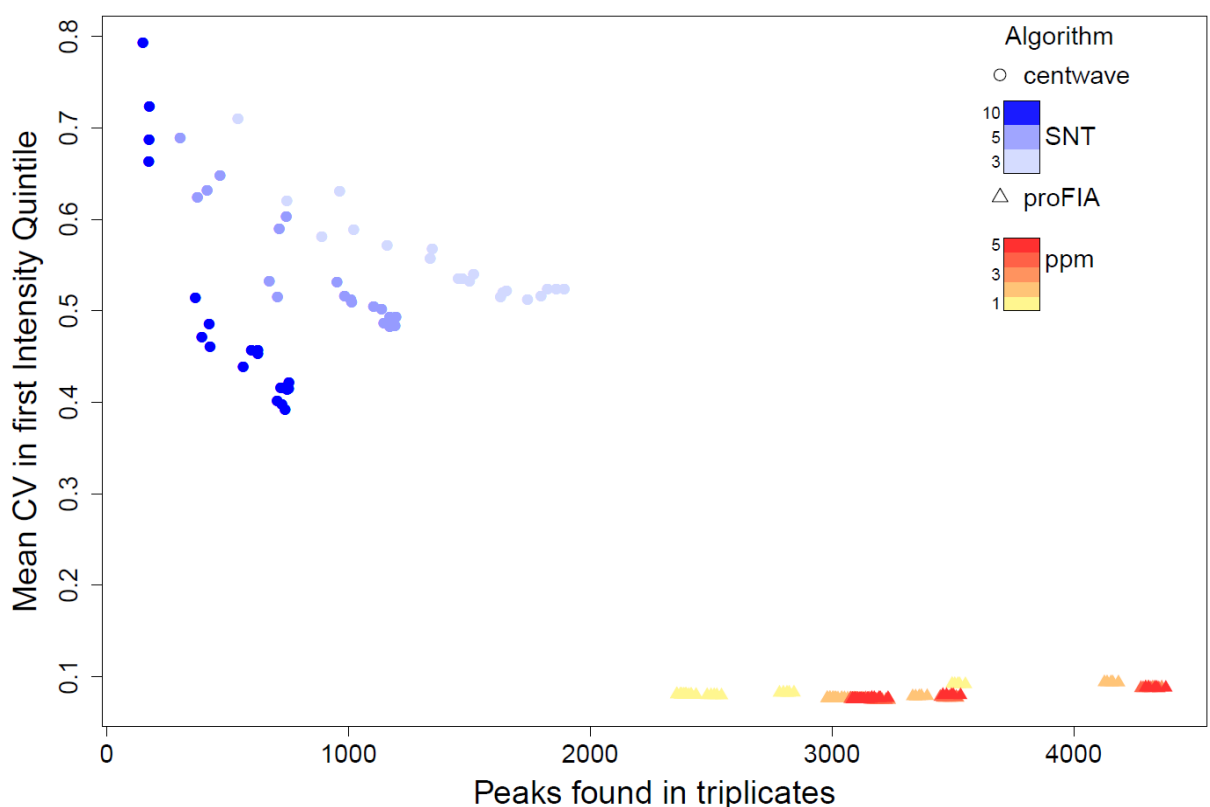

**Comparisons of *centWave* and *proFIA* for FIA-HRMS data preprocessing:** Each point represents a combination of parameters. Points are colored according to the values of the most significant parameter (*SNT* for *centWave* and *ppm* for *proFIA*).

*proFIA* outperforms *centWave* whatever the set of parameters used, which may be explained by:
- Detection efficiency: matrix effect in FIA may result in missing centroids within the mass trace, which are not allowed by *centWave*
- Accurate quantification: the second derivative Gaussian wavelet from *centWave* fails to detect FIA signal boundaries in case of peak shape alteration by matrix effect

Importantly, the good CV results from *proFIA* are stable whatever the parameterization used. Such a strength of the FIA algorithm is due to the fact that the sample peak model is estimated by using the most intense features, and is therefore robust to changes in parameter values.

The details of the influence of the *ppm*, *bandCoverage*, and *pvalthresh* parameters on *proFIA* peak detection and quantification are shown below: the following graphics display a zoom in of the previous figure colored according to each of the parameters.

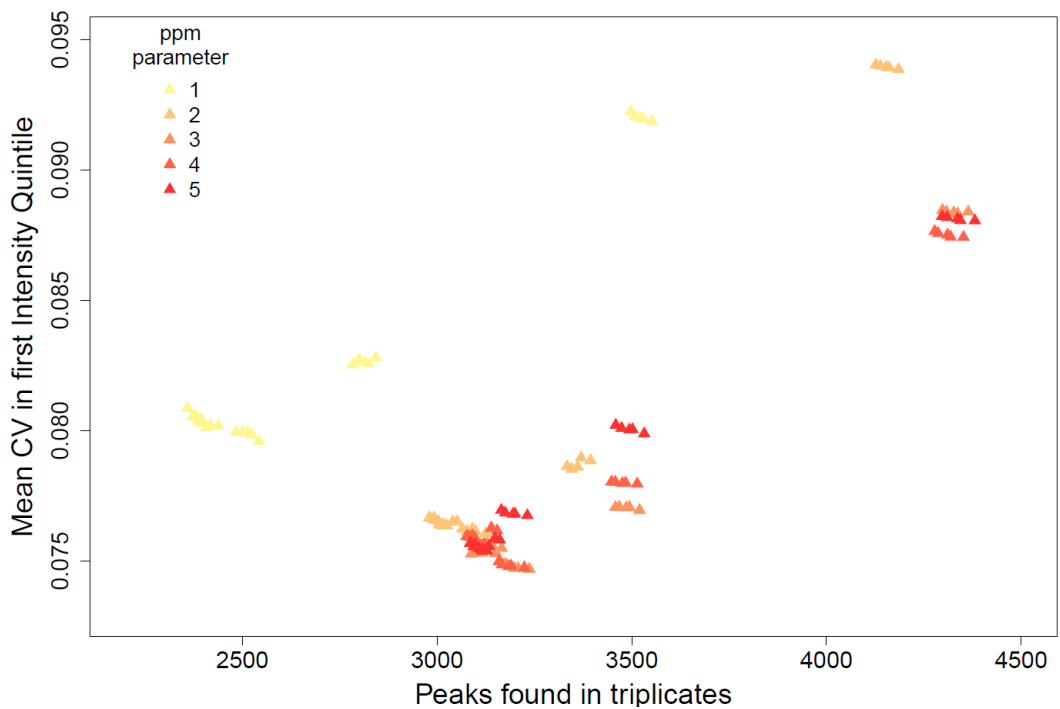

**Influence of the *ppm* parameter value on detection by *proFIA*.** The *ppm* parameter depends on the mass accuracy of the spectrometer. It is similar to the corresponding parameter from *centWave*. *ppm* was found to be the most discriminating parameter for the preprocessing of the *Fusion* dataset. Here, we see that whereas a 1 ppm value decreases the reproducibility of peak detection, the 2, 3, 4 and 5 ppm values give similar results.

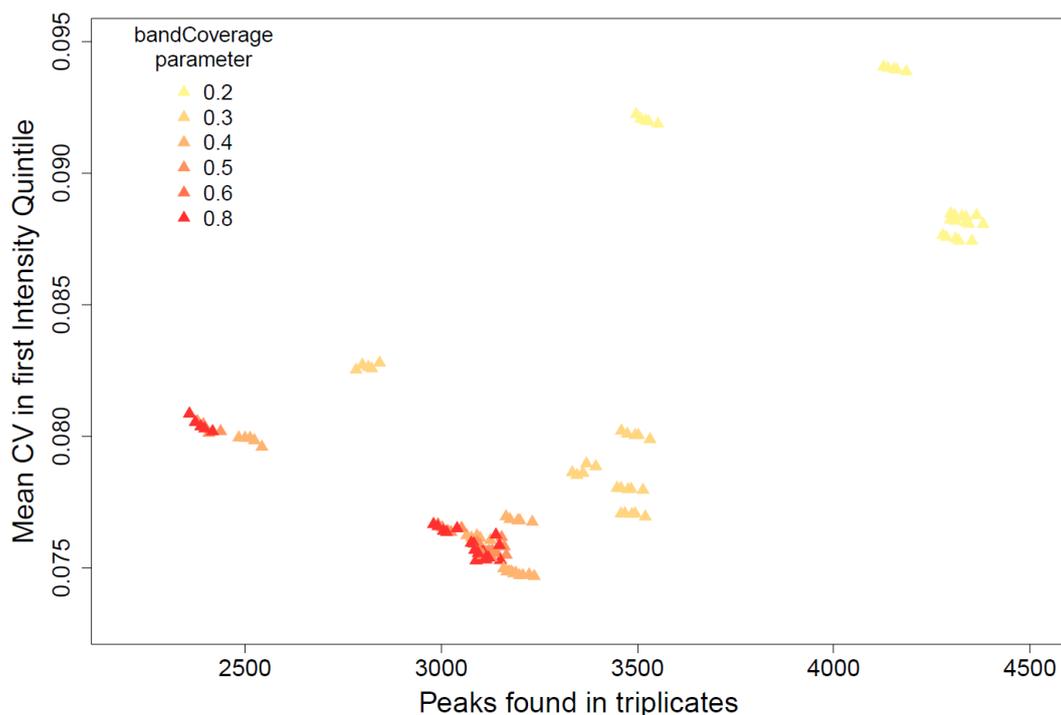

**Influence of the *bandCoverage* parameter.** *bandCoverage* is a filter used after band detection: only bands with more than a *bandCoverage* percent of centroids in the sample injection time window (or with more than *sizeMin* consecutive centroids, to also include signals with solvent baseline) are kept. The *bandCoverage* optimal value can vary depending on the acquisition. Here, values above 0.2 lead to a decrease of detection reproducibility. Values lower than 0.2 might also be considered to detect the signals with the strongest matrix effects.

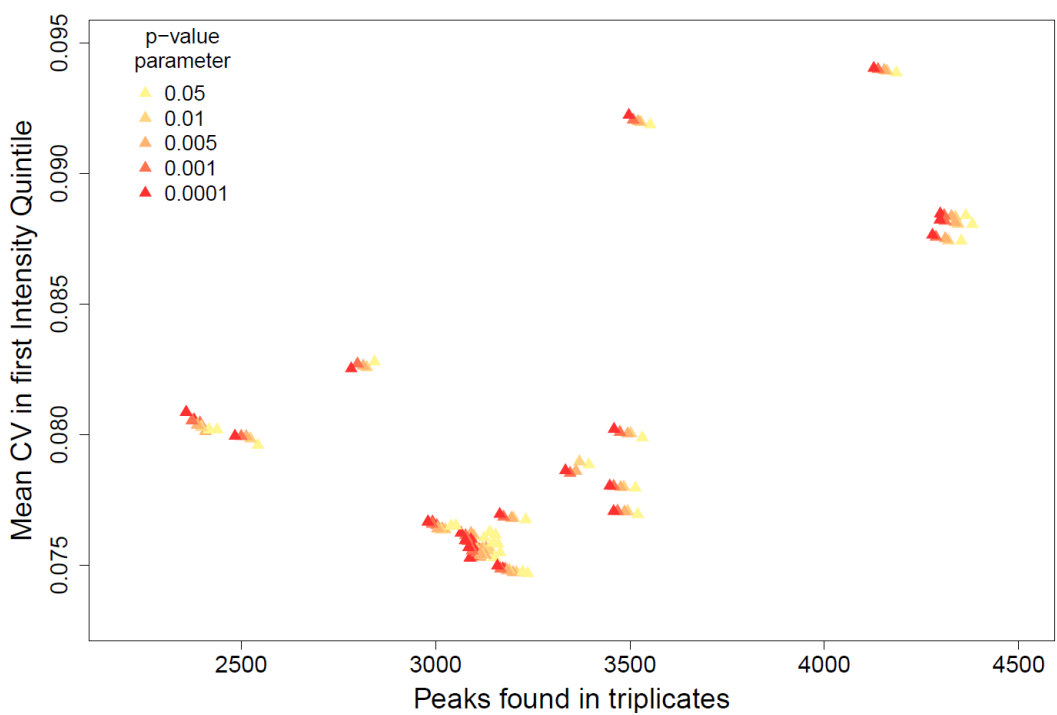

**Influence of the *pvalthresh* parameter.** *pvalthresh* is the threshold used to filter out the features whose intensity is not significantly higher than the solvent baseline. Since most of the bands containing solvent have been discarded during the band detection step, the *pvalthresh* parameter has little influence on the reproducibility of signal detection and quantification.

# Multivariate analysis of the generated peak table (*Fusion* dataset)

Multivariate modeling of the the spiking concentration was performed by using Partial Least Squares (*ropls* R/Bioconductor package[3]; see the figure below). As expected, samples with increasing concentration were aligned along the first component of the score plot (A). In addition, the spiked compounds had high loading values for this component (B). By coloring the loading plot according to our indicator of matrix effect (cor. samp. peak: computed as the mean correlation between the feature signal in each sample and the sample peak model), we observed that the indicator values were high for the spiked compounds (B and C). This may be explained by the fact that the matrix effect is negligible when compared to the concentration increase. Indicator values below 0.2 (threshold used in the *proFIA* diagnostic *plot()* method) tend to indicate a strong peak shape alteration by matrix effect in several samples (D).

---

[3] Thevenot, E.A.; Roux, A.; Xu, Y.; Ezan, E.; Junot, C. Analysis of the human adult urinary metabolome variations with age, body mass index and gender by implementing a comprehensive workflow for univariate and OPLS statistical analyses. *Journal Proteome Research,* **2015**, *14* (8), 3322-3335.

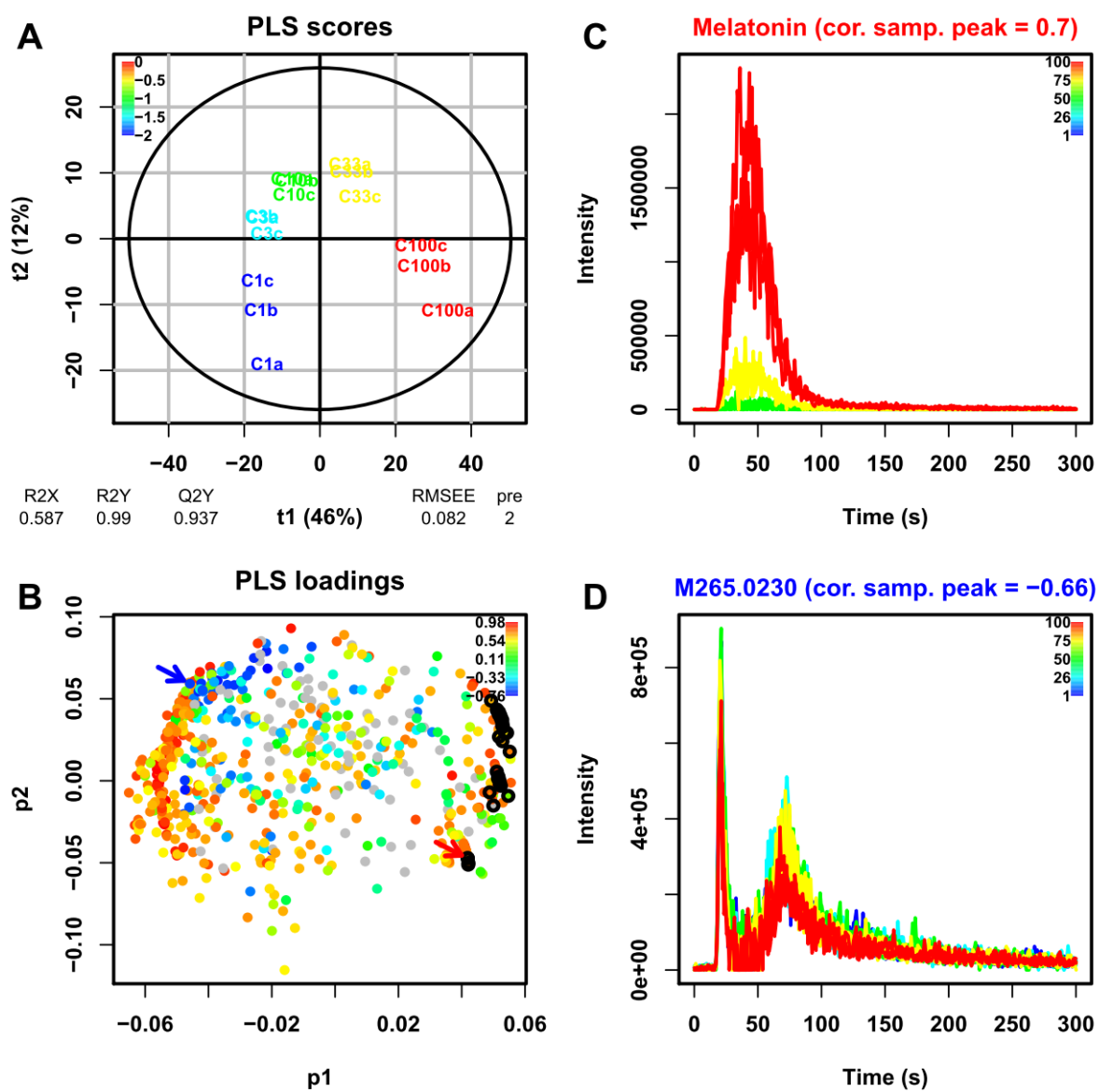

**Multivariate analysis of the FIA-HRMS data set following *proFIA* preprocessing.** The 15 raw files corresponding to the spiked samples were preprocessed with *proFIA*. The peak table was restricted to the features which were present in more than half of the samples, and the intensities were log10 transformed. (A) and (B): The spiking concentration was modeled by Partial Least Squares. The Q2Y value was significant, as assessed by random permutation of the response values ($p < 0.001$). On the score plot (A), C1 corresponds to the lowest spiking concentration (C1 = 10 ng/mL). On the loading plot (B), features are colored according to the matrix effect indicator (correlation with the sample peak). In cases where a time shift is detected in one of the feature flowgrams (e.g. because of retention in the injection pipe), the correlation value is no more relevant, and it is therefore set to NA (grey dots). The black circles correspond to the spiked compounds. (C) and (D): Flowgrams of two features (indicated by arrows on the loading plot) with high and low ME indicator values are displayed with the *plotFlowgrams()* method.